\documentclass[twocolumn,showpacs,preprintnumbers,amsmath,amssymb]{revtex4}

\usepackage{graphicx}
\usepackage{dcolumn}
\usepackage{bm}
\usepackage{latexsym}

\begin{document}

\title{Two-Poton Correlations of Luminescence under Bose-Einstein
Condensation of Dipolar Excitons}

\author{A.\,V.\,Gorbunov, V.\,B.\,Timofeev, D.\,A.\,Demin$^+$, A.\,A.\,Dremin}
\address{Institute of Solid State Physics, Russian Academy of Sciences, Chernogolovka, Moscow region,
142432 Russia}
\thanks{e-mail: gorbunov@issp.ac.ru}
\address{$^+$ Moscow Institute of Physics and Technology, Dolgoprudny, Moscow region, 141700 Russia}

\date{\today}

\begin{abstract}

Correlations of luminescence intensity (a 2$^{nd}$ order
correlator $g^{(2)}(\tau )$, where $\tau $ is the delay time
between photons in registered photon pairs) have been studied
under Bose-Einstein condensation of dipolar excitons in the
temperature range of 0.45$\div $4.2~K. Photoexcited dipolar
excitons were collected in a lateral trap in GaAs/AlGaAs
Schottky-diode heterostructure with single wide (25~nm) quantum
well under electric bias applied between heterolayers. Two-photon
correlations were measured with the use of a classical Hanbury
Brown-Twiss two-beam intensity interferometer with the time
resolution of $\approx 0.4$~ns. Photon ``bunching'' has been
observed near the Bose condensation threshold of dipolar excitons
that was determined by the appearance of a narrow luminescence
line of exciton condensate at optical pumping increase (FWHM of
the narrow line at the threshold $\lesssim 200$~$\mu $eV). The
two-photon correlation function shows super-poissonian
distribution, $g^{(2)}(\tau )> 1$, at time scales of system
coherence ($\tau _{c} \lesssim 1$~ns). No photon bunching was
observed with the used time resolution at the excitation pumping
appreciably below the condensation threshold. At excitation
pumping well above the threshold, when the narrow line of exciton
condensate begins to grow in the luminescence spectrum, the photon
bunching is decreasing and finally vanishes with further
excitation power increase. In this pumping range, the photon
correlation distribution becomes poissonian reflecting the
single-quantum-state origin of excitonic Bose condensate. Under
the same conditions a first-order spatial correlator,
$g^{(1)}(r)$, measured by means of the luminescence amplitude
interference from spatially separated condensate parts under cw
photoexcitation, remains significant on spatial scales around
4~$\mu $m. The discovered effect of photon bunching is rather
temperature-sensitive: it drops several times with temperature
increase from 0.45~K up to 4.2~K. If we assume that the
luminescence of dipolar excitons collected in the lateral trap
reflects directly coherent properties of interacting exciton gas,
the observed phenomenon of photon bunching nearby condensation
threshold -- where exciton density and hence luminescence
intensity fluctuations are most essential -- manifests phase
transition in interacting exciton Bose gas. It can be used as an
independent tool for exciton Bose condensation detection.

\end{abstract}

\pacs{73.21.Fg, 78.67.De}

\maketitle

1. Investigations of two-particle spatially-temporal correlations
(a 2$^{nd}$ order correlator $g^{(2)}(r,\tau ))$ are increasingly
popular particularly in connection with the analysis of complex
quantum collective phenomena in ensembles of ultracold atoms
\cite{Schellekens2005, Jeltes2007, Ottl2005}. The study of
radiation intensity correlations, or two-photon correlations, goes
back to a classical pioneering work of Hanbury Brown and Twiss
\cite{HBT} and its quantum grounds given later by Glauber
\cite{GLB}. It was demonstrated in these papers that photons
emitted by a chaotic light source tend to bunch only under
incoherent mixing or under superposition of coherent states (i.e.
correlator $g^{(2)}(\tau )$ shows super-poissonian distribution of
two-photon correlations at a mutual coherent time $\tau _{c}$ of a
radiation system). At the same time, in the case of a
single-quantum-state source, which is coherent in all orders (i.e.
single-mode laser \cite{GLB}, atom laser \cite{Ottl2005} or Bose
condensate of atoms \cite{Schellekens2005}), the correlator
$g^{(2)}(r,\tau )$ is exactly of poissonian type, with no bunching
effect. The behavior of a spatially-temporal correlator is defined
by quantum statistics of identical particles. Effect of bunching
appears only for bosons (photons, magnons, atoms-bosons, etc.),
whereas for fermions the two-particle correlations should exhibit
untibunching behavior, which was recently observed in experiments
for ultracold fermion atoms -- $^{3}$He \cite{Jeltes2007}.

Here we present results of our experimental study of luminescence
intensity correlations under conditions of Bose-Einstein
condensation of dipolar excitons. As far as we know, it is the
first experiment of this type for interacting dipolar exciton Bose
gas collected in a lateral trap. The importance of luminescence
intensity correlations exploration for exciton Bose condensation
had been pointed out earlier in the theoretical paper \cite{LHT}.
Recently, behavior of two-photon correlations was investigated
experimentally for 2D excitonic polaritons in microcavity
heterostructures with quantum wells \cite{KSPRZ1, KRIZH}.

2. Spatially-indirect or dipolar excitons were investigated in a
wide (25 nm) single GaAs quantum well (QW) placed in a
perpendicular to heterolayers electric field applied between a
metal film (Schottky gate) on a surface of
AlGaAs/GaAs-heterostructure and a built-in conductive electron
layer in a structure \cite{JETPL83, JETPL84}. Due to the applied
electric field, dipolar excitons have a big dipole moment in the
ground state (more than 100 D). In the system under investigation
such excitons do not bind into molecules or other multiparticle
complexes due to dipole-dipole repulsion.

Photoexcitation of excitons and luminescence registration was
performed through the circular window of {\o}5~$\mu $m size in a
metal mask (Schottky gate). Dipolar excitons were collected in a
ring lateral trap which arose along the window perimeter because
of strongly non-uniform electric field \cite{SUG, DAN}. An
increased image of the window was projected on an entrance slit of
a spectrometer supplied with a cooled silicon CCD camera. The
spectrometer transferred the image from the entrance slit plane to
the exit slit plane without aberrations (``imaging
spectrometer''), thus permitting to register sample image in a
zero order of diffraction grating. The utilized optical system
allowed to observe  a spatial structure of dipolar exciton
luminescence within the window in a metal gate with a resolution
down to 1~$\mu $m. The sample was mounted in a helium optical
cryostat with the working temperature range $T$ from 0.45~K up to
4.2~K. At $T < 1.5$~K, the sample was immersed directly into
liquid $^{3}$He, while at $T > 1.5$~K it remained in cold $^{3}$He
vapor. A spatial structure of luminescence could be studied with
spectral selectivity with the use of optical interference filters.
The optical scheme enabled not only to observe a luminescence
picture in a 5-$\mu $m window with a high spatial resolution, but
to carry out also -- by means of minor realignments -- an optical
Fourier-transform of images.

Dipolar excitons were excited by two cw lasers simultaneously:
with wavelength $\lambda = 782$~nm (photoexcitation with photon
energy less than the forbidden gap in AlGaAs barrier --
``under-barrier'' excitation) and $\lambda = 659$~nm
(``above-barrier'' photoexcitation). By adjusting the power ratio
between these lasers, we reached maximal compensation of extra
charges in the trap, and the system of excitons was maintained as
neutral as possible \cite{SOLOV, JETPL84}. The details of
architecture of the used structures, lateral traps and
compensation of extra charges in traps are given in \cite{JETPL83,
JETPL84, JAP, PSS}.

\begin{figure}[ptb]
\centerline{\includegraphics[width=.53\textwidth]{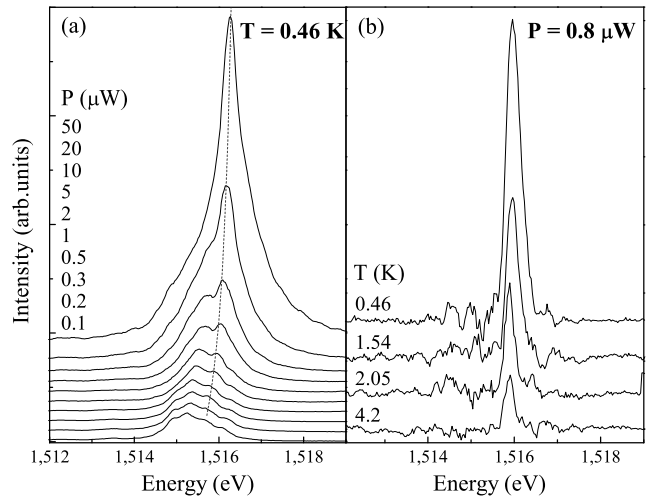}}
\caption{Photoluminescence spectra of dipolar excitons in a ring
lateral trap. (a) Threshold formation and increase of a narrow
spectral line with rising above-barrier laser power $P$ ($\lambda
 = 659$~nm). Power of under-barrier laser is fixed: $P_{782nm}
=10$~$\mu $W. Temperature $T = 0.46$~K (b) Narrow spectral line
growth (the background related to localized states is subtracted)
with temperature diminishing from 4.2~K to 0.46~K at fixed
photoexcitation: $P_{659nm} = 0.8$~$\mu $W, $P_{782nm} = 10$~$\mu
$W.} \label{fig1}
\end{figure}

3. Photon bunching was likely to be observed close to the
threshold of Bose condensation of dipolar excitons, where
fluctuations of exciton density are the strongest. Therefore, at
the beginning, the phase diagram of exciton Bose condensation in a
lateral trap was explored to define the equilibrium phase boundary
outlining an area of condensation in coordinates ``pumping power
(or exciton density) $P$ -- temperature $T$''. With this aim, the
luminescence spectra were investigated and analyzed at variation
of optical pumping in the temperature range of 0.45$\div $4.2~K.
On reaching conditions critical for condensation (both temperature
and pumping rate), a narrow line of dipolar excitons starts to
grow in luminescence spectrum. This event corresponds to
macroscopic filling of a lowest state in a trap and to occurrence
of exciton condensate \cite{PSS}. Fig.\ref{fig1}a demonstrates
spectra of dipolar exciton luminescence detected directly from a
ring trap under variation of cw optical pumping at $T = 0.46$~K.
In this case the pumping was carried out by both lasers with
$\lambda = 659$~nm and $\lambda = 782$~nm for maximal compensation
of extra charges in a trap, whereas only the power of
above-barrier photoexcitation ($\lambda = 659$~nm) was varied.

At very weak photoexcitation, an unstructured and asymmetric
luminescence band with FWHM about 1.3~meV is visible in a
spectrum. The shape of this band does not vary with pumping
reduction. The band is inhomogeneously broadened. It originates
from excitons localized on fluctuations of random potential
because of residual charged impurities and structural defects both
in the trap and in its vicinity. With excitation power increase at
the violet edge of the band, a narrow spectral line corresponding
to the condensed part of dipolar excitons appears above a certain
power threshold and grows further in intensity \cite{JETPL84,
PSS}. The linewidth near threshold is about 200~$\mu $eV, and its
intensity increases in this pumping range superlinearly. The
dependence of line intensity on pumping power becomes linear at
the further increase of excitation power. At high photoexcitation,
the line dominates in spectrum in comparison with an unstructured
underlying continuum. The line slightly broadens and moves towards
higher energies with pumping. Such a behavior is connected with a
repulsive interaction of dipolar excitons on their concentration
increase and it was analyzed in detail in \cite{ZIMM, STERN}.
According to our measurements, the spectral line shift and
broadening are comparable. In particular, in the range of
photoexcitation corresponding to Fig.\ref{fig1}a, the ratio of the
spectral shift of the line center of gravity (spectral moment
$M_{1})$ to its width (spectral moment $M_{2})$ is $M_{1}/M_{2}
\approx 1.25$. The maximal concentration of excitons in the
pumping range of Fig.\ref{fig1}a can be estimated using the
spectral shift of the line ($\lesssim $ 300~$\mu $eV): it does not
exceed $10^{10}$~cm$^{-2}$.

An intensity of the luminescence line corresponding to exciton
condensate is very sensitive to temperature. At fixed
photoexcitation power intensity decreases linearly with
temperature, down to its full disappearance in unstructured
continuum near the threshold of exciton condensation. The
temperature behavior of the narrow line of exciton condensate is
illustrated by Fig.\ref{fig1}b. In the temperature range, $T =
0.45\div 4.2$~K the following law for the narrow line intensity
$I(T)$ at fixed pumping was established: $I(T) \sim (1 -
T/T_{c})$, where $T_{c}$ is a critical temperature and the narrow
spectral line disappears above $T_{c}$.

\begin{figure}[ptb]
\centerline{\includegraphics[width=.52\textwidth]{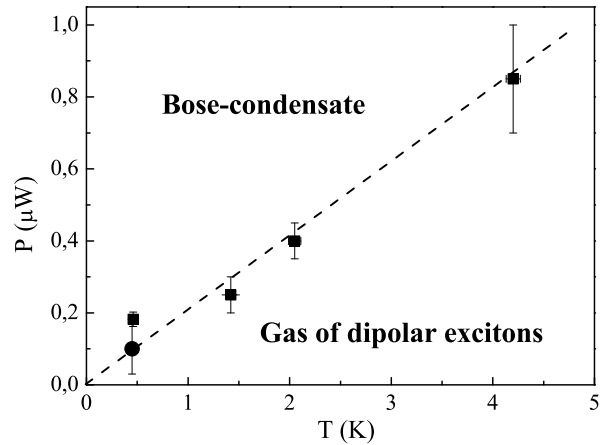}}
\caption{The phase diagram of the Bose condensation of dipolar
excitons in a ring trap in coordinates ``above-barrier
photoexcitation power $P_{659nm}$ -- temperature $T$''.
Under-barrier photoexcitation is fixed: $P_{782nm} =10$~$\mu $W.
Squares correspond to the condensation threshold determined by the
appearance of a narrow spectral line. The circle at $T = 0.45$~K
shows the threshold estimated from the maximum position of
two-photon correlator vs pumping.} \label{fig2}
\end{figure}

To plot phase diagram at each fixed temperature in the explored
interval $T = 0.45\div 4.2$~K, the dependence of luminescence
spectra on photoexcitation power was investigated. As a result,
the threshold power $P_{thr}$ was defined when the narrow line of
exciton condensate started to appear (disappear) in spectrum. The
phase diagram was plotted in coordinates $P - T$, and for its
construction the nonlinear range of the narrow line intensity
dependence on laser excitation power was used. Fig.\ref{fig2}
represents the resultant phase diagram. The phase boundary
outlining the area of Bose condensation looks like a linear
function of temperature, as it is expected for two-dimensional
system.

The previous measurements of such phase diagram were performed for
dipolar excitons in a structure with double QW \cite{DREMIN}. A
large concentration of structural defects in such a structure and,
as a consequence, too high exciton mobility edge prevented from
correct measurements at low density of free excitons at $T < 1$~K.
In the present structure with single wide QW, the density of
structural defects is almost ten times as low. In this case one
manages to deal with free excitons, i.e. above mobility edge, down
to 0.45~K. As a result, the linear extrapolation of phase boundary
to the area of even lower temperatures and smaller exciton
densities (see Fig.\ref{fig2}) directly approaches the origin of
coordinates.

4. Simultaneously with the narrow line origination in luminescence
spectra of the Bose condensate of dipolar excitons, a
spatially-periodic pattern of equidistantly located luminescent
spots appears in the luminescence image, which is observed
directly from a 5~$\mu $m-window with spatial resolution about
1~$\mu $m and with spectral selection by means of optical
interference filter with a bandwidth of 11~{\AA} (details are
presented in \cite{JETPL84, JAP, PSS}. The structure of
luminescence spots at fixed pumping also turned out
temperature-sensitive: the spatially-periodic structure smears
into a continuous ring at $T \lesssim 10$~K. \textit{In situ}
optical Fourier transforms of the spatially-periodic patterns
which reproduce luminescence intensity distribution in far zone,
showed results of both destructive and constructive interference,
and also a spatial directivity of the luminescence normal to
heterolayers. These result from large-scale coherence of the
condensed exciton state in a lateral ring trap. Direct
measurements of two-beam interference from pairs of
spatially-separated luminescence spots in a ring allowed to
estimate the length of spatial coherence, and also the value of
the amplitude correlator $g ^{(1)}(r) \backsimeq 0.2$ at a
distance not less than 4~$\mu $m. This large scale of spatial
coherence means that the experimentally observed periodic
luminescent structures are described under the conditions of Bose
condensation of dipolar excitons in lateral trap by a single wave
function. It should be emphasized that large spatial coherence is
exhibited both by Bose condensate of exciton polaritons in
microcavity heterostructures with several QWs \cite{KSPRZ2} and by
a collective state of spatially indirect excitons in structures
with double QWs \cite{YANG}.

5. Now let us turn to the study of two-photon correlations of
luminescence intensity under the conditions of exciton Bose
condensation. We measured a correlator of intensities:
\[
g^{(2)}(\tau )=\frac{<I_1 (r,t)I_2 (r,t+\tau >}{<I_1 (r,t)><I_2 (r,t>}
\]
Here angular brackets mean ensemble averaging, $r$ is a spatial
coordinate of emitter, and $\tau $ is a delay time between photons
in a registered pair.

\begin{figure}[ptb]
\center{\includegraphics[width=.55\textwidth]{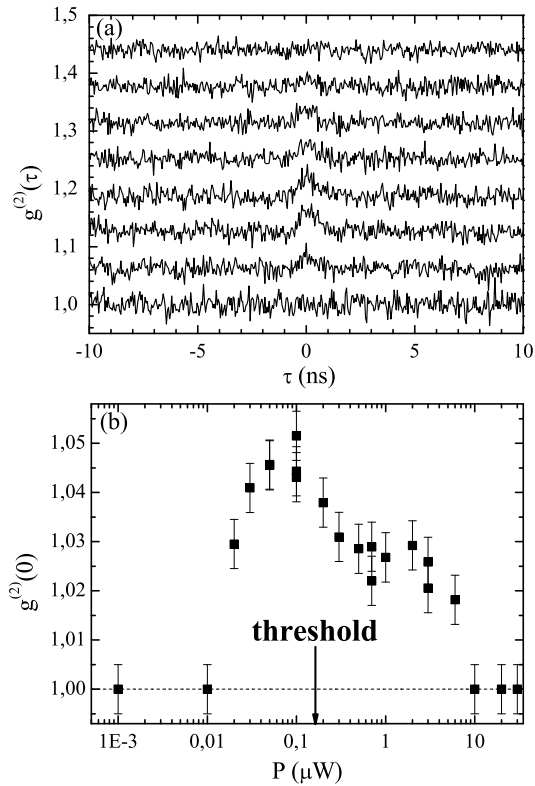}}
\caption{Second-order correlator $g^{(2)}$\textit{($\tau )$} for
the luminescence of dipolar excitons in a ring trap as a function
of photoexcitation $P$. (a) Time diagrams $g ^{(2)}$\textit{($\tau
)$} at power $P_{659nm}$ = 0.01, 0.02, 0.05, 0.1, 0.3, 2, 6 and
30~$\mu $W from bottom to top, respectively. The graphs are
shifted vertically for convenience. (b) The value of $g^{(2)}(0)$
vs power. The arrow shows the Bose condensation threshold in
accordance with the phase diagram in Fig.\ref{fig2}. $P_{782nm}
=10$~$\mu $W. $T = 0.45$~K.} \label{fig3}
\end{figure}

Measurements of two-photon correlations were carried out with the
use of a two-beam interferometer of intensities according to the
classical scheme of R.~Hanbury Brown and R.~Q.~Twiss \cite{HBT}.
High-speed avalanche photodiodes (Perkin-Elmer SPCM-AQR-16) with
rise time of $\approx400$~ps were used as single-photon detectors.
They were placed symmetrically with respect to non-polarizing
beam-splitting cube which equally divides a luminescence light
beam coming from the sample. Special screens and diaphragms
provided registration only of the useful signal of luminescence
and completely excluded effects of diffused light and difficult to
control reflections. Photoresponse signal arrived at the start
trigger of the electronic "time-amplitude" converter (ORTEC, TAC
567) from one detector and at the stop trigger -- from another.
Its output was connected to the input of multichannel analyzer
(ORTEC, TRUMP-PCI-8K). The technique allowed to detect the
intensity correlator $g^{(2)}(\tau )>1$ under the conditions of
superposition of coherent states with mutual coherence time
$\backsimeq0.5$~ns. Correlation measurements of luminescence
intensity were carried out within a narrow spectral band
containing the line of exciton condensate. This band was cut out
from the luminescence spectrum (presented in Fig.\ref{fig1}à) by
an interference filter with full spectral bandwidth of
$\approx2$~meV. However, this spectral selection could not get rid
of the unstructured luminescence background located directly under
the narrow line of exciton condensate.

\begin{figure}[ptb]
\center{\includegraphics[width=.48\textwidth]{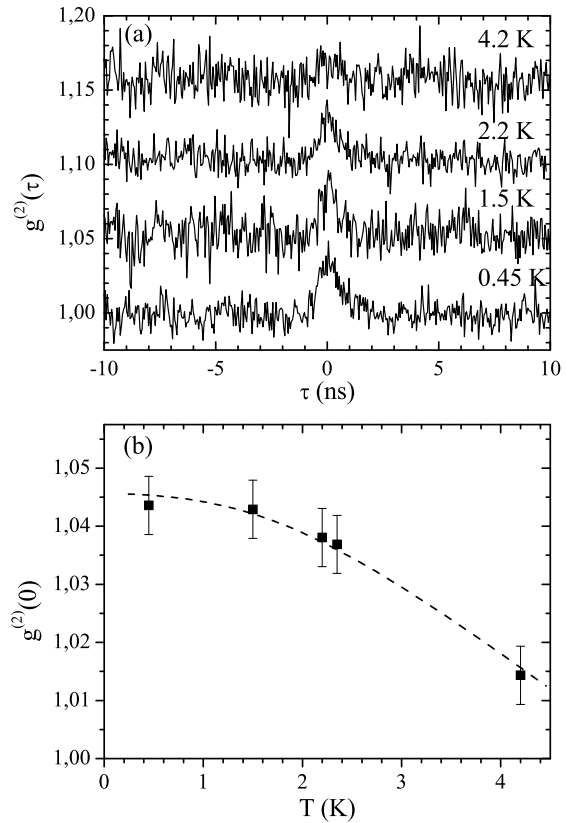}}
\caption{Second-order correlator $g^{(2)}(\tau )$ for the
luminescence of dipolar excitons in a ring trap as a function of
temperature $T$. (a) Time diagrams $g^{(2)}(\tau )$. The graphs
are shifted vertically for convenience. (b) The value of
$g^{(2)}(0)$ vs $T$. $P_{659nm} = 0.1$~$\mu $W, $P_{782nm}
=10$~$\mu $W.} \label{fig4}
\end{figure}

The results are presented in Figs.\ref{fig3} and \ref{fig4}. We
will consider measurements at $T = 0.45$~K (Fig.\ref{fig3}) in
detail. At pumping rate considerably below condensation threshold
the distribution of two-photon correlations is poissonian. In this
range of photoexcitation only unstructured spectrum of
luminescence as wide as 1.3~meV is observed, that corresponds to
localized exciton states. Once again we emphasize that the time
resolution of the used registration system does not allow us to
detect superposition of chaotic coherent sources (in our case
these are the localized exciton states) if their times of
coherence are much less than 0.4~ns. When approaching condensation
threshold, the effect of photon bunching appears and grows with
further increase of optical pumping. Thus, the function of
two-photon correlations shows super-poissonian distribution, $g
^{(2)}(\tau )>1$, on coherence time scales of the investigated
system $\tau \lesssim 1$~ns. The measured value of photon bunching
is limited by two factors. The former is due to the fact that
beside the luminescence of dipolar excitons, an input of the wide
spectral continuum caused by localized states is also recorded.
The latter is related to the finite time resolution of the used
registration system. At pumping appreciably exceeding the
threshold, when the narrow line of exciton condensate dominates in
luminescence spectra, the effect of bunching decreases and
finally, completely vanishes with further increase of optical
excitation. The distribution of two-photon correlations becomes
poissonian, thus reflecting, as we assume, a single quantum
coherent state of exciton Bose condensate. This conclusion proves
to be true by direct observations of a large-scale coherence
(\textit{in situ} Fourier images) and by a 1$^{st}$-order
correlator (two-beam interference from spatially separated spots
in a luminescence pattern).

\begin{figure}[ptb]
\centerline{\includegraphics[width=.52\textwidth]{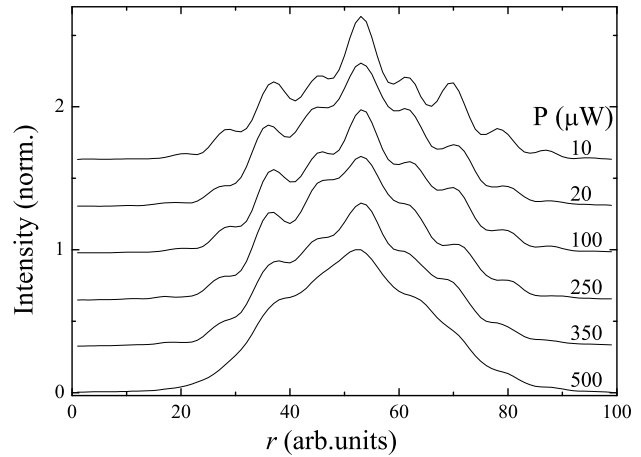}}
\caption{Amplitude interference at the luminescence light
overlapping from two spatially separated ($\approx4$~$\mu $m)
condensate parts (two diametrically opposite spots in the
luminescence pattern of dipolar excitons in a ring trap) at
different photoexcitation powers $P$ ($\lambda = 633$~nm). The
first-order correlator $ g^{(1)}(r) \approx 0.2$ for $P = 10$~$\mu
$W and $g^{(1)}(r)\to 0$ at $P = 500$~$\mu $W. The graphs are
shifted vertically for convenience. $T = 1.7$~K.} \label{fig5}
\end{figure}

Fig.\ref{fig5} illustrates the behavior of the interference
pattern measured by means of the luminescence amplitude
interference from two spatially separated ($\approx4$~$\mu $m)
condensate parts under variation of photoexcitation power. In the
pumping range, where two-photon correlator approaches unity,
$g^{(2)}(\tau ) \to 1$, i.e. where the photon bunching disappears
completely, the measured 1$^{st}$ order spatial correlator is
rather significant, $g^{(1)}(r) \approx 0.2$. Only at very high
photoexcitation, almost ten times higher than the largest used in
the whole set of data presented in Figs.\ref{fig1}-\ref{fig4}, the
interference pattern in Fig.\ref{fig5} starts to wash out. In this
high photoexcitation range, the exciton Bose condensate is
destroyed due to increasing phase decoherence processes
(presumably because of strong dipole-dipole repulsive exciton
interaction).

The observed effect of photon bunching proved to be very sensitive
to temperature. The measured magnitude of photon bunching
decreases several times on temperature increase over the range
$0.45 \div 4.2$~K (see Fig.\ref{fig4}). This observation may
indirectly indicate destruction of an order parameter with
temperature. We also emphasize that the regions of maximal photon
bunching measured at various temperatures vs photoexcitation power
correlate well with the found phase diagram (Fig.\ref{fig2}). It
means that the maximal two-photon bunching occurs in the area
where fluctuations of exciton density are the strongest, i.e.
close to the phase boundary. Under the same experimental
conditions, no bunching is observed in the spectral range of
direct exciton luminescence.

Having assumed that the luminescence of dipolar excitons directly
reflects coherent properties of interacting exciton gas, the
observed phenomenon of photon bunching nearby condensation
threshold, where fluctuations of exciton density and hence
luminescence-intensity fluctuations are most significant,
manifests phase transition in the interacting exciton Bose gas and
it can be used as an independent tool to detect exciton Bose
condensation. It must be interesting to further investigate under
Bose condensation of dipolar excitons the spatial correlator of
luminescence intensity, $g^{(2)}(r)$, which is directly connected
with the off-diagonal order parameter. Besides, in the
photoexcitation range, where the 1$^{st}$ order correlator
evidently exhibits the effect of decoherence, $g^{(1)}(r)\to 0$,
due to disintegrating dipolar exciton Bose condensate, two-photon
correlations should be studied at much higher time resolution.

Authors would like to thank L.V.Keldysh for the interesting
discussions of presented results. The work is supported by Russian
Foundation for Basic Research, the Presidium of Russian Academy of
Sciences (programme on nanostructures) and the Branch of Physical
Sciences of Russian Academy of Sciences (programme on strongly
correlated systems).

\end{document}